\documentclass[journal]{IEEEtran}
\usepackage{cite}
\usepackage{graphicx}
\usepackage{caption}
\usepackage{subcaption}
\usepackage{fullpage}
\usepackage{authblk}
\usepackage{xcolor}
\usepackage{amsmath}
\usepackage{breqn}
\usepackage{lipsum}
\usepackage{mathtools, cuted}

\newcommand*{\Scale}[2][4]{\scalebox{#1}{$#2$}}%

\begin{document}
\title{From Fowler-Nordheim to Non-Equilibrium Green's Function Modeling of Tunneling}
\author[1*]{Hesameddin Ilatikhemeneh}

\author[2*]{Ramon B. Salazar}

\author[1]{Gerhard Klimeck}

\author[1]{Rajib Rahman}

\author[2]{Joerg Appenzeller}

\affil[1]{Network for Computational Nanotechnology, Department of Electrical and Computer Engineering, Purdue University, West Lafayette, IN 47907, USA}

\affil[2]{Birck Nanotechnology Center, Department of Electrical and Computer Engineering, Purdue University, West Lafayette, IN 47907, USA}
\affil[*]{\normalsize{These authors contributed equally to this work.}}
\maketitle
\section{Abstract}
In this work, an analytic model is proposed  which provides in a continuous manner the current-voltage characteristic (I-V) of \emph{high performance} tunneling field-effect transistors (TFETs) based on direct bandgap semiconductors. The model provides closed-form expressions for I-V based on: 1) a modified version of the well-known Fowler-Nordheim (FN) formula (in the ON-state), and 2) an equation which describes the OFF-state performance while providing continuity at the ON/OFF threshold by means of a term introduced as the \emph{"continuity factor"}. It is shown that traditional approaches such as FN are accurate in TFETs only through correct evaluation of the total band bending distance and the \emph{"tunneling effective mass"}. General expressions for these two key parameters are provided. Moreover, it is demonstrated that the tunneling effective mass captures both the ellipticity of evanescent states and the dual (electron/hole) behavior of the tunneling carriers, and it is further shown that such a concept is even applicable to semiconductors with nontrivial energy dispersion. Ultimately, it is found that the I-V characteristics obtained by using this model are in close agreement with state-of-the-art quantum transport simulations both in the ON- and OFF-state, thus providing validation of the analytic approach. 

\section{Introduction}
Due to the increasing role of band-to-band tunneling (BTBT) and Schottky barrier (SB) tunneling mechanisms in electronic devices, it becomes critical to develop analytic models that allow identifying the key parameters which optimize the tunneling performance. One of the most promising tunneling device concepts is the tunneling field-effect-transistor (TFET) \cite{Appenzeller1, Appenzeller2, Wenjun} whose performance can be significantly boosted by aggressively scaling down its dimensions (referred to as \emph{high performance} TFETs) \cite{Energy_TFET, Hesam1}. The importance of tunneling is not limited to the tunneling devices but it also affects the performance of ultra-scaled FETs significantly \cite{Sub12}.

Many approaches to TFET modeling are based on the numerical evaluation of the Wentzel-Kramers-Brillouin (WKB) approximation \cite{WKB3,WKB4,WKB5,WKB6,Analytic1}. The WKB method has been found to be an appropriate description for  TFETs based on direct bandgap semiconductors with a single dominant tunneling path if the WKB inputs, namely analytic potential profile and complex bandstructure, are close to those obtained from self-consistent atomistic calculations \cite{WKB6, Analytic1}. In other works the WKB approximation has been solved analytically for a triangular potential barrier which leads to the traditional Fowler-Nordheim (FN) tunneling transmission expression \cite{FN,WKB1,WKB2}. 
\begin{equation}
\label{eq:FN1}
T_{WKB}^{FN}=exp\left(-\frac{4}{qF}  \frac{\sqrt{2m_t^*} E_g^{3/2}}{3\hbar} \right) \\
\end{equation}
\noindent where $q$ is the elementary charge, $\hbar$ is the reduced Planck constant, $F$ is the electric field in the transition region from source to channel, $E_g$ is the bandgap of the semiconductor, $m_t^*$ is the \emph{"tunneling effective mass"}. 

It is commonly assumed that $m^*_t$=$m^*_r$, where $m^*_r$ is the so called \emph{"reduced effective mass"} \cite{Kane}. This approach has been used in previous TFET modeling efforts \cite{WKB3,WKB1,WKB2,E,F}. Assuming $m^*_t$=$m^*_r$ implies that the complex bands within the bandgap of the semiconductor are parabolic. In other works, $m^*_t$ is assumed to be either the conductivity effective mass \cite{WKB5,H}, or is simply treated as a fitting parameter \cite{mass_tune,J}. In this study it is shown that the appropriate definition of $m^*_t$ allows a more realistic description of 1) the ellipticity of the complex bandstructure inside the bandgap (not parabolic) and, 2)  the electron/hole duality of tunneling carriers, which have both shown to be key components in modeling the BTBT process \cite{elliptic}. 

In addition, the correct evaluation of the \emph{"total band bending distance ($\Lambda$)"}, another key parameter in the FN formula, is discussed for different types of junctions. The use of $\Lambda$ and $m_t^*$ leads to a modified FN formula which improves its accuracy in the calculation of tunneling transmission in the ON-state of TFETs.

A new analytic model is ultimately presented based on 1) a modified version of the well-known Fowler-Nordheim (FN) equation for the on-state, 2) a new OFF-current formula which accounts for the direct tunneling from source to drain, while providing a smooth ON-OFF transition through the introduction of a \emph{"continuity factor"}. The combination of these analytic approaches allows describing the current-voltage characteristics (I-V) both in the ON- and OFF-state of the device by use of simple and closed-form expressions. The results provided by the analytic model are in close agreement with state-of-the-art full-band numerical simulations using the Non-Equilibrium Green’s Function (NEGF) formalism.

\section{Tunneling effective mass}
\label{mt}
\begin{figure}[!t]
        \centering
        \begin{subfigure}[b]{0.24\textwidth}
                \includegraphics[width=\textwidth]{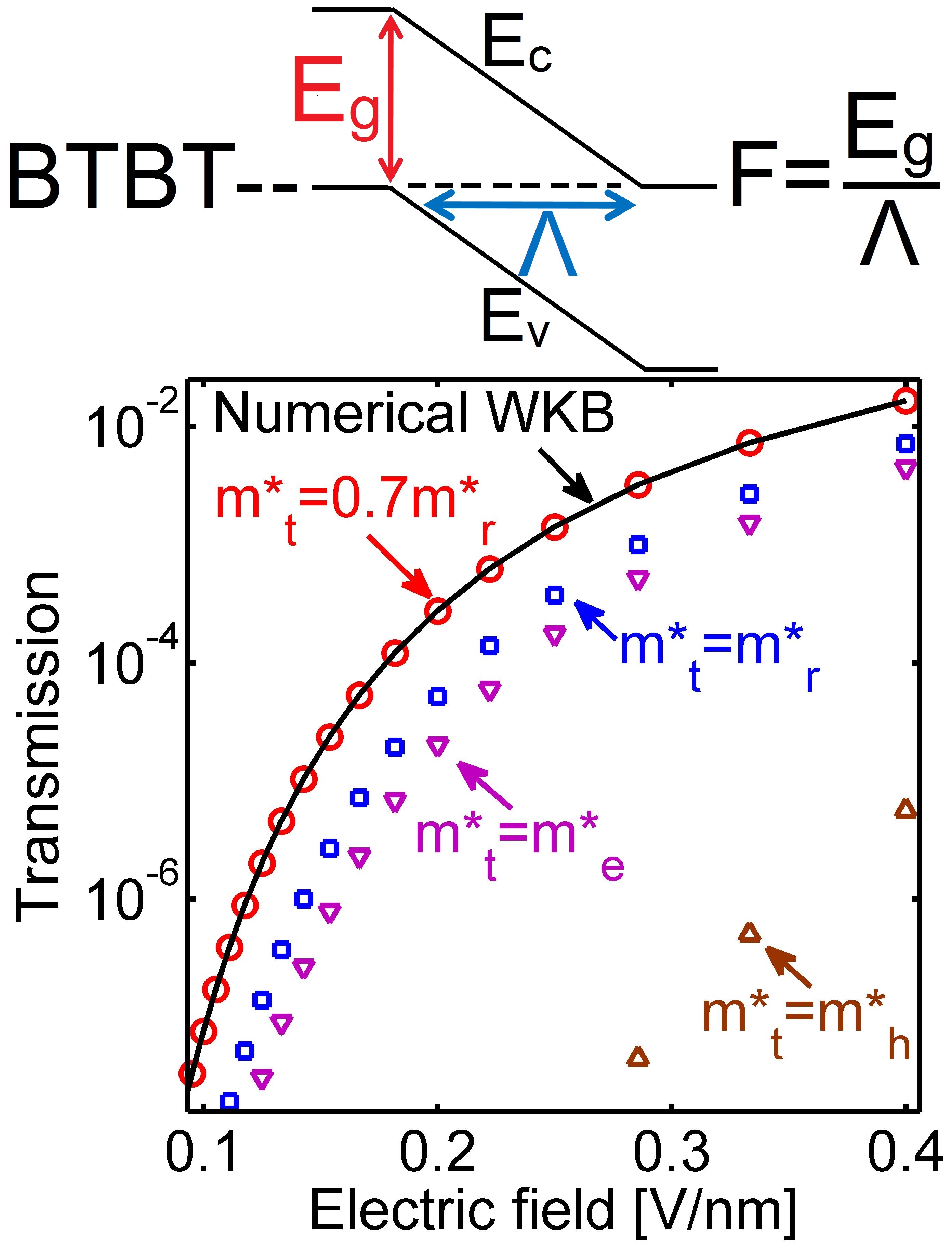}
                \vspace{-.8\baselineskip}
                \caption{}
                \label{fig:IdVg_dop}
        \end{subfigure}%
        ~ 
        \begin{subfigure}[b]{0.26\textwidth}
                \includegraphics[width=\textwidth]{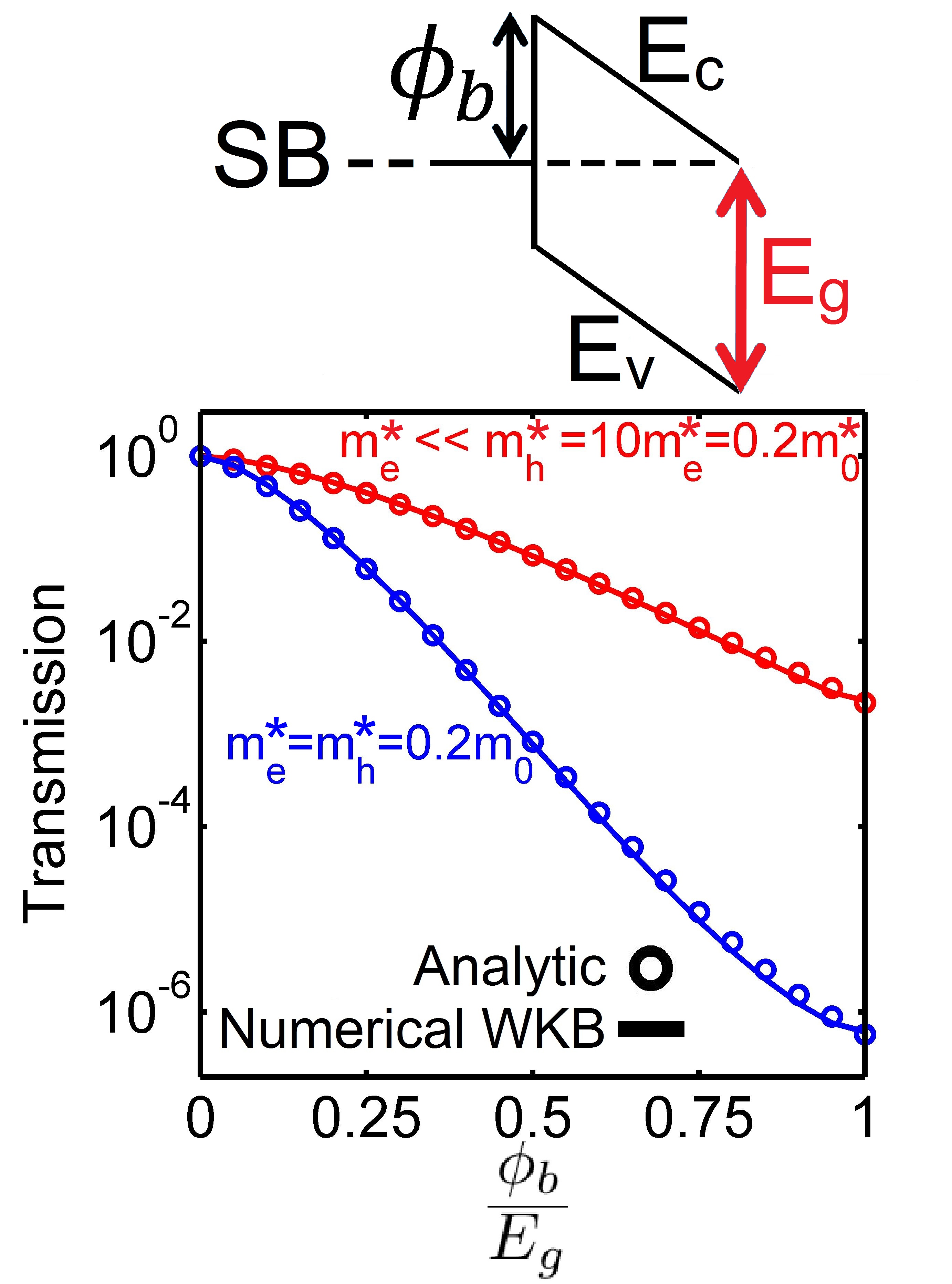}
                \vspace{-.8\baselineskip}
                \caption{}
                \label{fig:Pot_dop}
        \end{subfigure}
                     \vspace{-1.2\baselineskip}
        ~ 
        \caption{a) BTBT process with an electric field of $F$ (top) and BTBT transmission for a material with $m_h^*=5m_e^*=0.5m_0$ and $E_g=1eV$ using numerical WKB (black line), $m_t^*=0.7 m_r^*$ (red circles), $m_t^*=m_r^*$ (blue squares), $m_t^*=m_e^*$ (pink triangles), and $m_t^*=m_h^*$ (brown triangles). b) A Schotttky barrier with a barrier height of $\phi_b$ (top) and its tunneling transmission probability with $m_e^*=m_h^*$ (blue) and $m_e^* \ll m_h^*$ (red) using analytic equation (\ref{eq:mt_gen2}) (circle symbols) and a numerical evaluation of the WKB integral (lines). }\label{fig:Fig1}
\end{figure}

Previously, it has been demonstrated that the complex part of the band-structure is an elliptic curve (not parabolic) which is composed of two parts: one dominated by hole behavior, and another by electron behavior. Such an elliptic curve can be described by an analytic formula within 1.4\% error compared to rigorous tight-binding simulations \cite{elliptic}. It can be mathematically proven that by taking into account the ellipticity of the complex bands and by assuming a constant electric field in the tunneling window, the tunneling transmission is given by (see Appendix \ref{App_A} for details)
\begin{equation}
\label{eq:Elp}
T_{WKB}^{Elliptic}=exp\left(-\frac{\pi}{qF}  \frac{\sqrt{m_r^*} E_g^{3/2}}{2\hbar} \right) \\
\end{equation}
\noindent where $m_r^*$ is the reduced effective mass:
\begin{equation}
\label{eq:mr}
m_r^* = \frac{m_e^* m_h^*} {m_e^* + m_h^*} \\
\end{equation}
Note that (\ref{eq:Elp}) is almost identical to the transmission probability equation derived in the 1D case by Kane except for a prefactor of $(\pi/3)^2$ \cite{Kane, Kane2}. Therefore, the Kane's approach takes into account the elliptic complex band structure precisely even for materials with $m^*_e \neq m^*_h$. Through direct comparison, one can realize that equations (\ref{eq:FN1}) and (\ref{eq:Elp}) are equivalent expressions if the following condition is satisfied:
\begin{equation}
\label{eq:mt}
m_t^*|_{BTBT} ~= \left(\frac{3\pi}{8 \sqrt{2}} \right)^2 m_r^* \approx 0.7 m_r^* \\
\end{equation}
In other words, equation (\ref{eq:FN1}) can capture the elliptic nature of the complex bandstructure if indeed $m_t^*$ from (\ref{eq:mt}) and NOT $m_r^*$ from (\ref{eq:mr}) is used. This also corrects the notion that the concept of effective mass cannot be applied to BTBT due to the dual electron/hole behavior of carriers as they tunnel through the bandgap \cite{duality}. For illustration purposes, Fig. \ref{fig:Fig1}a compares the BTBT transmission obtained using different choices of $m_t^*$ for a material with $m^*_h$=$5m^*_e$=$0.5m_0$ and $E_g$=$1eV$. The results obtained using $m^*_t$=$0.7 m^*_r$ matches well with the numerical WKB calculation, whereas the other conventional choices of $m_t^*$  (i.e. $m_t^*$=$m_r^*$ or $m_t^*$=$m_e^*$) lead to a significant error.

To further elucidate the practical advantage and general applicability of the concept of the tunneling effective mass, we consider two examples: 1) graphene nanoribbons (GNRs) with non-trivial energy dispersion and 2) metal-semiconductor Schottky barriers (SBs).
It is well established that for GNRs, the band edge effective mass is given by $m^*_{GNR} = \frac{E_g}{2v_f^2}$ \cite{Datta}, where $v_f$ is the group velocity of carriers within the energy range in which the dispersion is linear. By combining (\ref{eq:FN1}), (\ref{eq:mr}) and (\ref{eq:mt}) and using $m_h^*$ = $m_e^*$ = $m^*_{GNR}$ due to the symmetry in the energy dispersion, the expression shown in equation (\ref{eq:GNR}) is obtained. It is noteworthy that (\ref{eq:GNR}) matches exactly the result reported in previous studies but found through more convoluted arguments \cite{duality, jena}.            
\begin{equation}
\label{eq:GNR}
T_{WKB}^{GNR} = exp\left(-\frac{\pi}{qF}  \frac{E_g^{2}}{4\hbar v_f} \right) \\
\end{equation}
The equation for $m_t^*$ can also be generalized to metal-semiconductor junctions where the injected carrier tunnels only through a portion of the band gap (${\phi_b}/{E_g}$) as shown in Fig. \ref{fig:Fig1}b (see Appendix \ref{App_B} for details). 
\begin{equation}
\label{eq:mt_gen2}
m_t^*|_{SB} ~\approx 0.7 m_r^* \left( \frac{\phi_b}{E_g} \right) + m_{e}^* \left(1-\frac{\phi_b}{E_g} \right)\\
\end{equation}
Notice that (\ref{eq:mt_gen2}) should be used in the FN equation (\ref{eq:FN1}) using $\phi_b$ instead of $E_g$. Equation (\ref{eq:mt_gen2}) has a simple interpretation; when $\phi_b$ is small, tunneling occurs in the electron branch of the complex bandstructure, close to the conduction band edge, and hence $m_t^* \approx  m_e^*$, whereas for large $\phi_b$, $m_t^*$ reduces to the above reported BTBT result: $m_t^* \approx 0.7 m_r^*$. The analytic value of the tunneling transmission through a SB using (\ref{eq:FN1}) and (\ref{eq:mt_gen2}) is found to be in good agreement with the numerical evaluation of the transmission using the WKB approximation as shown in Fig. \ref{fig:Fig1}b for both cases of  $m_e^*=m_h^*$ (blue) and $m_e^* \ll m_h^*$ (red). 

\section{Junction electric field}
\begin{figure}[!b]
        \centering
        \begin{subfigure}[b]{0.5\textwidth}
                \includegraphics[width=\textwidth]{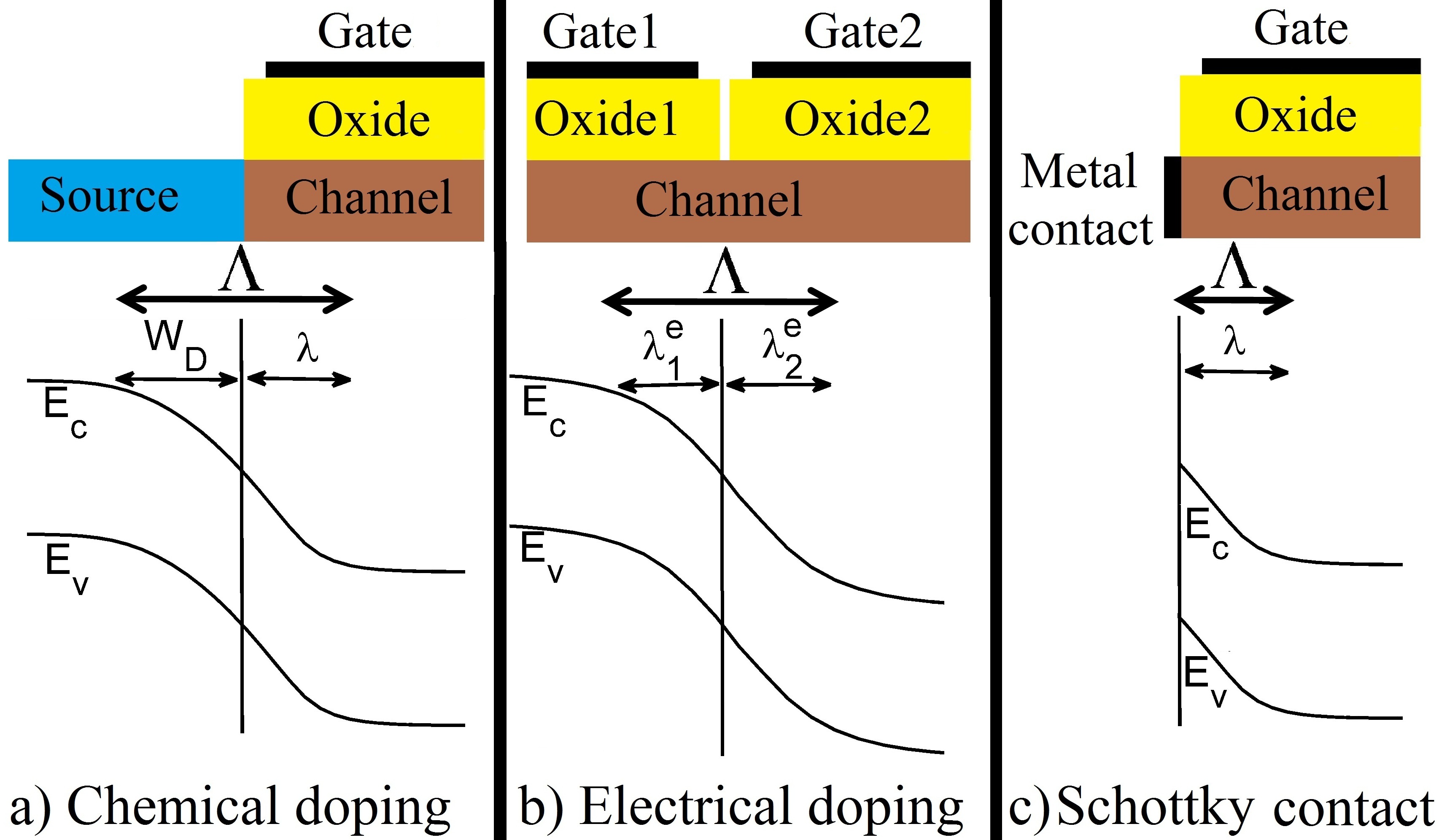}
                \label{fig:IdVg_dop}
        \vspace{-1.0\baselineskip}        
        \end{subfigure}
        \vspace{-0.5\baselineskip}        
        \caption{Total band bending distance for a device with a) chemically doped, b) electrically doped, and c) Schottky contact. }\label{fig:Lambda}
\end{figure}

 In addition to the tunneling effective mass, the electric field at the tunnel junction plays an important role in the tunneling transmission probability. The electric field magnitude depends on the junction configuration. i.e. the junction can be made in different ways: a gated channel in conjunction with a source/drain contact which is: 1) chemically  doped, 2) electrically doped, or 3)  a metal. The electric field can be quantified in terms of the source-to-channel band bending distance. Fig. \ref{fig:Lambda} shows the components of bending distance in different junction configurations (i.e. scaling length $\lambda$ and depletion width $W_D$).

Lee's scaling length theory \cite{Scaling1} can be used to evaluate the bending distance under the gate. According to theory \cite{Scaling1, Scaling2, Scaling3, Hesam2} the potential under the gate decays as $exp(-x/\lambda)$, where $\lambda$ is the "natural scaling length" and is known for different geometric architectures. For example, in cylindrical nanowire FETs, $\lambda$ is given by \cite{Scaling3}:
\begin{equation}
\label{eq:scaling2}
\lambda=\sqrt{  \frac{\epsilon_{ch}t_{ch}^2}{8\epsilon_{ox}}  \ln(1+ \frac{2t_{ox}}{t_{ch}} ) + \frac{t_{ch}^2}{16}} \\
\end{equation}
where $\epsilon_{ch}$, and $\epsilon_{ox}$ are the dielectric constants of the channel and the oxide respectively, while $t_{ch}$, and $t_{ox}$ are their thicknesses. 

Usually, the bending distance in the source/drain region is ignored. However, it was shown by us that for \emph{high performance} ultra-scaled TFETs the band bending distance in the source/drain contact is important and in many cases is dominant compared with $\lambda$ \cite{Analytic1}. The equations for the bending distance in different junction configurations shown in Fig. \ref{fig:Lambda} are:
\begin{equation}
\label{eq:Lambda}
\Lambda = 
\begin{cases}
W_D + \lambda & Chemically~doped~contact \\
\lambda_1^e + \lambda_2^e & Electrically~doped~contact \\
\lambda & Metal~contact \\
\end{cases}
\end{equation}

\noindent where $\Lambda$ is the total band bending distance and $W_D$ is the depletion width of the doped contact \cite{Analytic1}. Although different scaling theories are proposed for nanowire TFETs \cite{Scaling3, Taur}, the difference between $\lambda$ values is much smaller than $W_D$ even for highly doped TFETs \cite{Analytic1} and hence $\Lambda$ does not vary significantly. Notice that the equation for the scaling length is different in \emph{electrically doped} devices (distinguished by superscript $e$) compared to chemically doped devices \cite{Hesam2, Hesam3, 6GBLG}.
\begin{equation}
\label{eq:new_scaling5}
\lambda^e_{1/2} = \frac{T_{tot} + S/4}{\pi} 
\end{equation} 
where $T_{tot}$ is the total thickness of the device (e.g. $2t_{ox} + t_{ch}$ in the case of double-gated electrically doped TFETs) and $S$ is the spacing between the gates \cite{Hesam2}.
\section{Analytic current equation for TFET}

In this section, a model which provides the I-V of TFETs through a closed-form equation is presented. In the ON-state, this model uses the modified FN formula. Whereas in the OFF-state, the direct tunneling is combined with a term called the \emph{"continuity factor"} which ensures the continuous ON/OFF transition. Finally, the complete I-V model is presented based on the combination of the ON and OFF state analytic expressions.

\subsection{ON-state}
Considering the correct tunneling effective mass $m_t^*$ and the total bending distance $\Lambda$, the equation for current in the ON-state of the device can be written by combining the modified FN formula with the Landauer approach:
\begin{equation}
\label{eq:ON}
I=\frac{2q}{h} (\phi_s-\phi_{ch}-E_g)  exp\left(-\frac{4\Lambda}{\phi_s-\phi_{ch}}  \frac{\sqrt{2 m_t^* E_g} E_g}{3\hbar} \right) \\
\end{equation}
where $\phi_s$, and $\phi_{ch}$ are the electrostatic potentials at the source, and channel,  respectively. Notice that the difference in source and drain Fermi functions in the Landauer formula is assumed to be 1 ($f_S-f_D$=1) in the tunneling window (i.e. $\phi_s-\phi_{ch}-E_g$, see Fig. \ref{fig:off_2}). 
The expression shown in (\ref{eq:ON}) is applicable in the ON-state of the device where $\phi_s-\phi_{ch}>E_g + \delta$. The ON-OFF transition is slightly shifted by $\delta$ (defined in section \ref{res}) to have a continuous transition in I-V. 

\begin{figure}[!t]
        \centering
        \begin{subfigure}[b]{0.35\textwidth}
                \includegraphics[width=\textwidth]{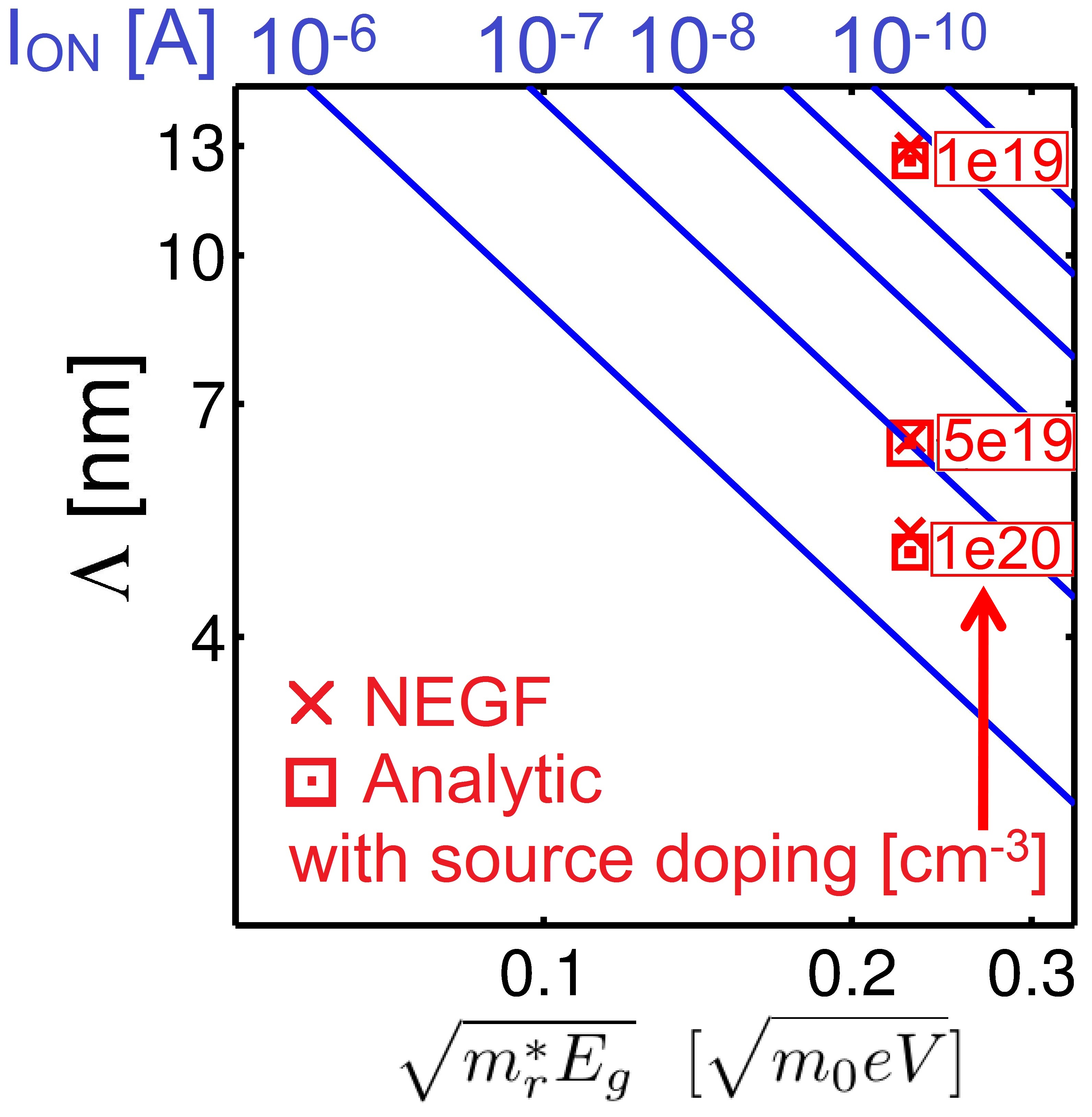} 
                        \vspace{-1.5\baselineskip}
                \label{fig:IdVg_dop}
        \end{subfigure}%
        \caption{Constant current lines as a function of the device design $\Lambda$ and channel material properties $\sqrt{m_{r}^*E_g}$ (blue lines) and $I_{ON}$ data points for InAs nanowire TFETs with a diameter of 3.4nm, oxide thickness of 1nm, $\epsilon_{ox}=9$, channel length of 15nm, $\Delta \approx 1/3$ and V$_{ds}$ of 0.5V obtained from full band atomistic simulations (cross symbols) and equation (\ref{eq:ON2}) (square symbols).}\label{fig:cte_I}
\end{figure}

The maximum achievable current ($I_{ON}$) in a TFET can be obtained by combining (\ref{eq:mt}) and (\ref{eq:ON}) and evaluating the result at $V_{gs} = V_{ds} = V_{DD}$ (i.e. $\phi_s-\phi_{ch} \approx E_g + qV_{DD} - qV_{th}$).
\begin{equation}
\label{eq:ON2}
I_{ON}= \Delta \frac{2q}{h} E_g ~ exp\left(- \frac{\pi}{2\hbar} \eta \right) \\
\end{equation}
\noindent where $\eta$ is the ON-current efficiency factor and $\Delta$ is the overdrive ratio defined as
\begin{equation}
\label{eq:eta}
\eta = \frac{ \Lambda \sqrt{m_{r}^*  E_g} } {1 + \Delta}
\end{equation}
\begin{equation}
\label{eq:Delta}
\Delta \approx \frac{q(V_{DD}-V_{th})}{E_g} 
\end{equation}
Equation (\ref{eq:eta}) shows how the device design (through $\Lambda$) and the channel material (through $E_g$ and $m_{r}^*$)  can impact the performance of the device. Fig. \ref{fig:cte_I} shows constant current lines (constant $\eta$) as a function of $\Lambda$ and $\sqrt{m_{r}^*  E_g}$. These constant current curves provide a practical guideline to assess the impact of the channel material and device design on the ON-current in a quantitative manner. In order to illustrate the accuracy and usefulness of this approach, the ON-currents of InAs nanowire TFETs with three different source doping levels are plotted in Fig. \ref{fig:cte_I}. Notice that for the same bias conditions ($\Delta \approx 1/3$) and the same type of device (InAs nanowire TFET) a change in source doping and accordingly in $\Lambda$ leads to several orders of magnitude of change in ON-current, which is found to be in good agreement with the NEGF simulations \cite{Hesam2}. 
\subsection{OFF-state} 
\begin{figure}[!t]
        \centering
        \begin{subfigure}[b]{0.24\textwidth}
                \includegraphics[width=\textwidth]{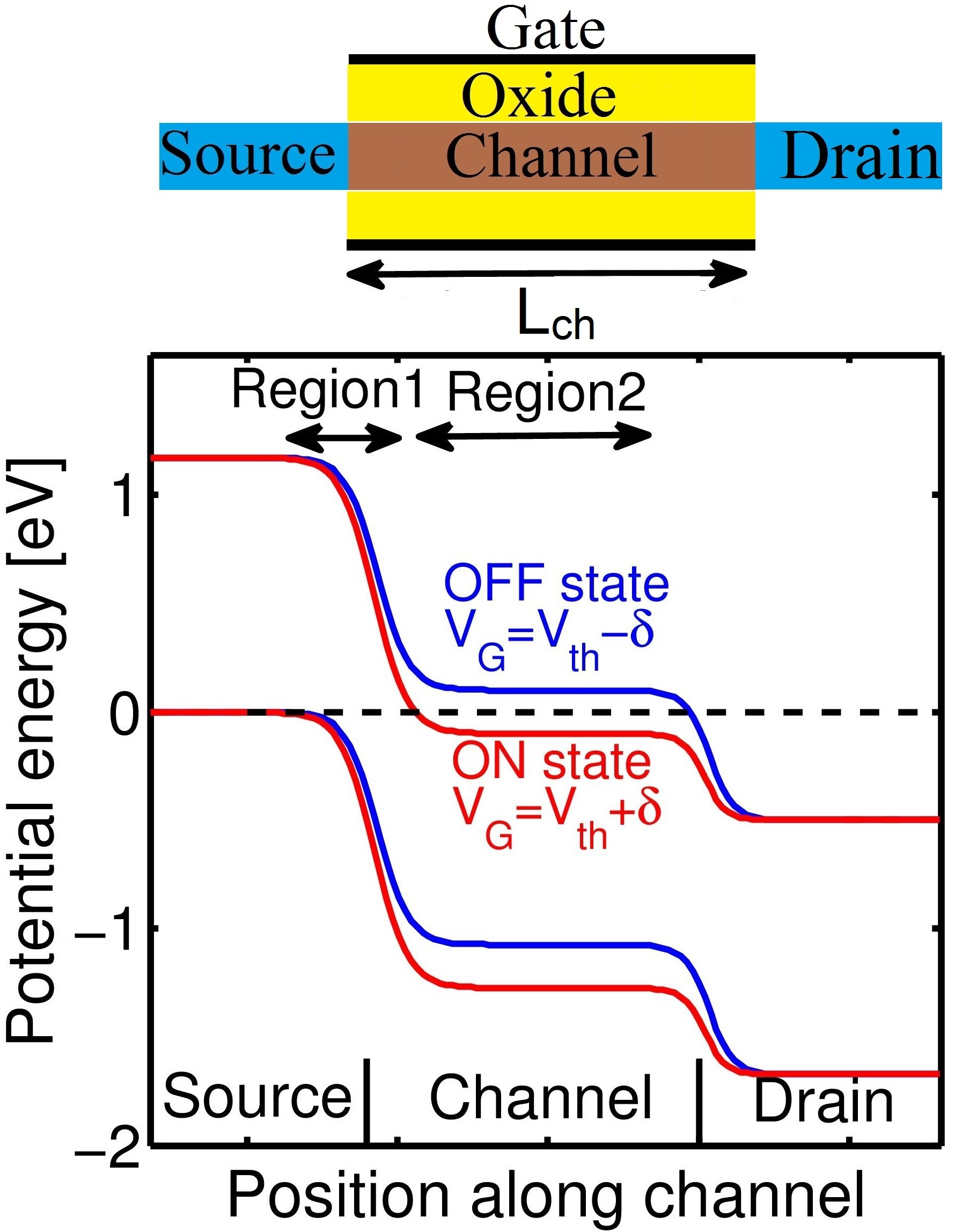}
   \vspace{-1.0\baselineskip}                
                \caption{}
                \label{fig:IdVg_dop}
        \end{subfigure}%
        \begin{subfigure}[b]{0.25\textwidth}
                \includegraphics[width=\textwidth]{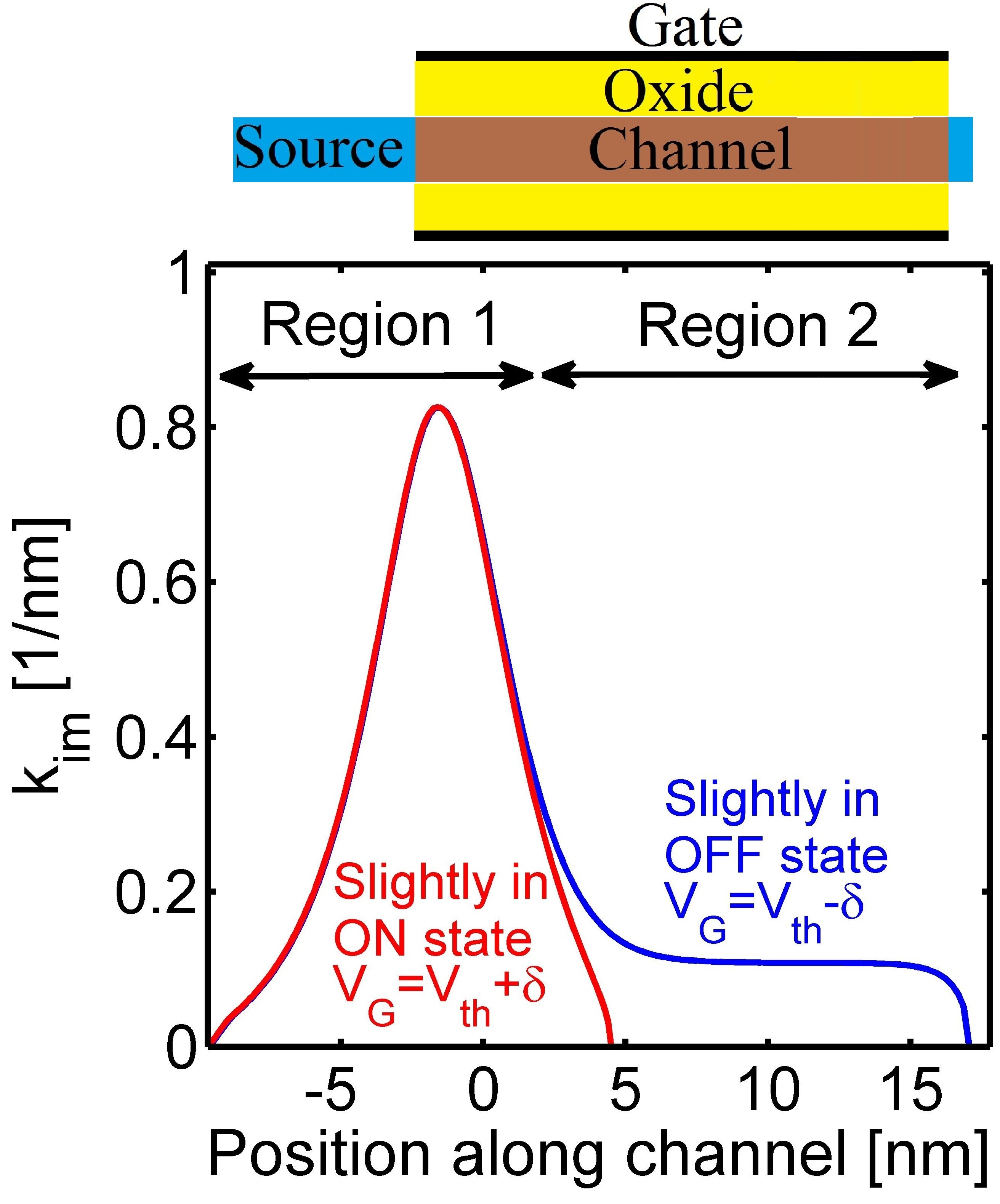}
                \caption{}
                \label{fig:Pot_dop}
        \end{subfigure}
      \vspace{-1.0\baselineskip}                
        ~ 
        \caption{a) Band diagram and b) imaginary wave-vector right before and after the threshold showing the importance of the continuity factor. The cross section of the structures are shown on top of the figures.}\label{fig:off2}
\end{figure}

The tunneling current in the OFF-state of the device ($I_{OFF}$) between source and drain in its general form is given by (see Fig. \ref{fig:off_2} in Appendix \ref{App_C})
\begin{equation}
\label{eq:OFF1}
I_{OFF}= \frac{2q}{h} \int_{\phi_d}^{\phi_s-Eg}T_{OFF}(E)dE
\end{equation}
The tunneling in the OFF-state occurs within 2 different regions: 1) a nearly triangular barrier at the source-channel interface and 2) a constant barrier with an approximate length equal to the channel length, as shown in Fig. \ref{fig:off2}a. 
\vspace{-1.2\baselineskip}
\begin{equation}
\label{eq:OFF2}
\vspace{-0.2\baselineskip}
T_{OFF}(E) = \underbrace{e^{-2 \int_{x_i}^{x_i + \Lambda} k_{im}(x)dx }}_{T_1}  \underbrace{e^{-2 \int_{x_i+\Lambda}^{x_i+\Lambda+L^'} k_{im}(x)dx } }_{T_2} 
\end{equation}
where $L^'$ is the effective channel length (i.e. $L_{ch}+W_{D}$ \cite{Analytic1}). Often, the first term $T_1$ (which is related to the crest shown in Fig. \ref{fig:off2}b) is ignored in the OFF-current calculations because in the case of long devices or bias conditions deep into the OFF-state, $T_2$ is much smaller than $T_1$ and $T_2$ determines the OFF-current. However, for aggressively scaled devices and for bias conditions close to threshold this term is quite important. This term also ensures current continuity at the ON/OFF threshold and thus is called the \emph{continuity factor $\gamma$}. The value of $\gamma$ can be found as follows: close to threshold the term $T_2 \approx 1$ because $k_{im} \approx 0$ (see Fig. \ref{fig:off2}a and \ref{fig:off2}b). In addition, using (\ref{eq:Elp}) close to threshold where $F \approx E_g/\Lambda$ leads to $T_1 \approx exp \left( - {\Lambda}  \frac{\pi}{2\hbar}  \sqrt{m_r^* E_g}\right)$. Deep into the OFF-state the term $T_2$ approaches $ e^{-2k_{im} (E)L^'}$ because $k_{im}$ is spatially constant throughout the channel. Therefore, the general form of the tunneling transmission in the OFF-state reads as:
\begin{equation}
\label{eq:OFF3}
T_{OFF}(E) \approx \underbrace{e^{- {\Lambda}  \frac{\pi}{2\hbar}  \sqrt{m_r^* E_g}}}_{\gamma} \times e^{-2k_{im} (E)L^'} \\
\end{equation}
Evaluating the integral over energy as in (\ref{eq:OFF1}), leads to (see Appendix \ref{App_C} for details)
\begin{equation}
\label{eq:OFF4}
\Scale[0.97]{
I_{OFF}=\frac{2q}{h}\gamma \left(\theta(|\frac{\phi_d-\phi_{ch}+E_g}{E_g}|)- \theta(|\frac{\phi_s-\phi_{ch}}{E_g}|) \right)}
\end{equation}

where $\phi_d$ is the electrostatic potential at the drain and the function $\theta$ is defined in Appendix \ref{App_C}. 

\section{Results and Comparisons} \label{res}
\begin{figure}[!t]
        \centering
        \begin{subfigure}[b]{0.4\textwidth}
                \includegraphics[width=\textwidth]{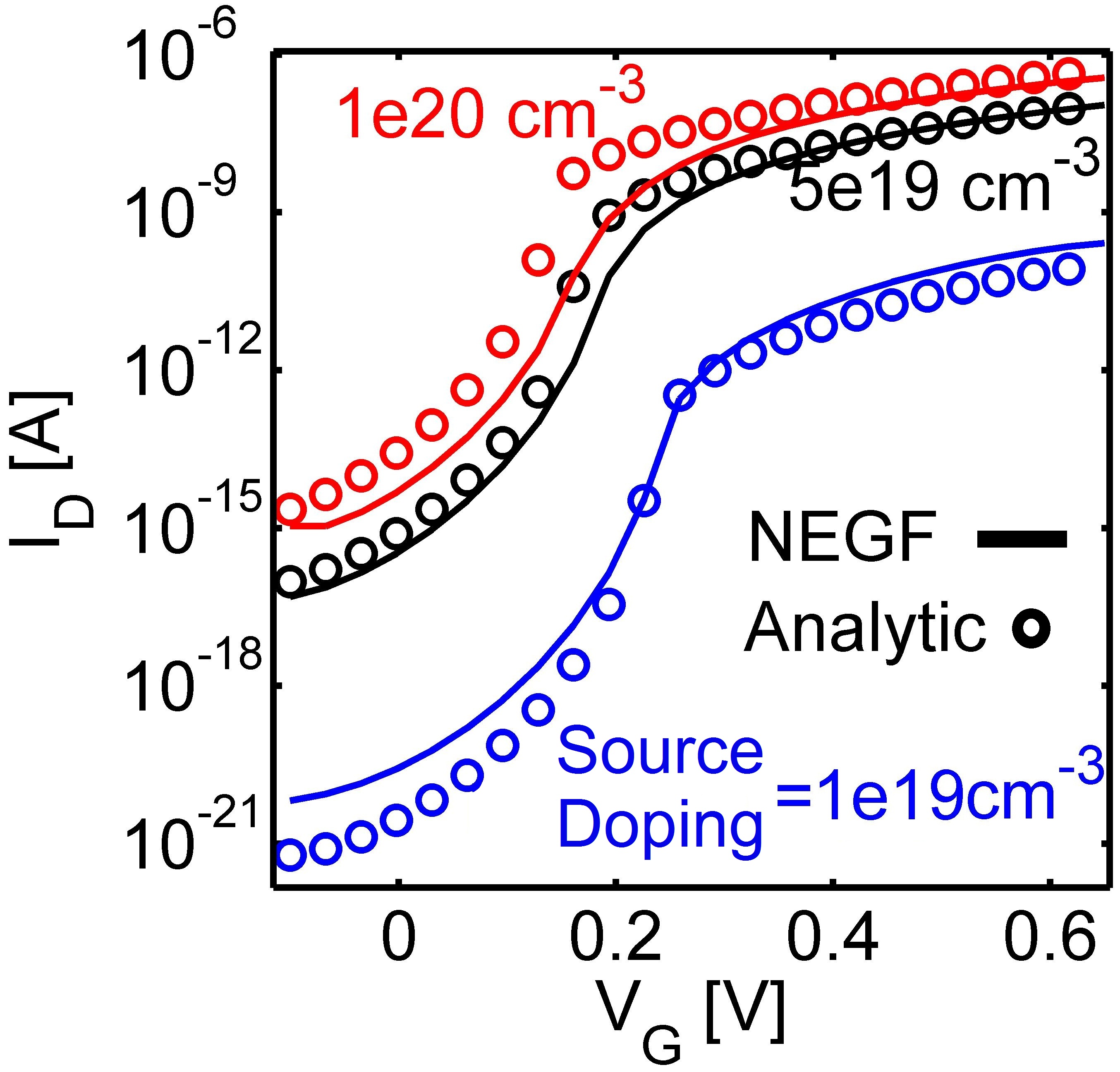}
                \label{fig:IdVg_dop}
        \end{subfigure}
        \caption{Transfer characteristics of the InAs nanowire TFETs with a diameter of 3.4nm, an oxide thickness of 1nm, $\epsilon_{ox}=9$, a channel length of 15nm, and V$_{ds}$ of 0.5V calculated from full band atomistic simulations (lines) and analytic equation \ref{eq:IV} (circle symbols) for different source/drain doping levels). }\label{fig:IV_supercompact}
\end{figure}

In order to validate the model, the results of the analytic approach are compared against the full-band quantum transport simulations of an ultra-scaled InAs nanowire TFET with a diameter of 3.4nm, a channel length of 15nm, and an oxide thickness of 1nm. The InAs nanowire is described by a $sp^3d^5s^*$ tight-binding model.  The numerical simulations are based on a self-consistent solution of the Poisson-NEGF equations using the nanodevice modeling tool NEMO5 \cite{nemo5_1, nemo5_2, nemo5_3}. Figure \ref{fig:IV_supercompact} shows the comparison of BTBT I-V (transfer) characteristic at $V_{ds}$ of 0.5V for three different source/drain doping values. The results have been obtained by NEGF (lines) and the analytic model (circle symbols) which is presented in equation (\ref{eq:IV}) given by the combination of (\ref{eq:ON}) and (\ref{eq:OFF4}) (ON- and OFF-state respectively): 
\begin{equation}
\label{eq:IV}
\Scale[0.75]{
I = 
\begin{cases}
\frac{2q}{h} (\phi_s-\phi_{ch}-\frac{E_g}{q})  exp\left(-\frac{4\Lambda \sqrt{2 m_t^* E_g} E_g}{3\hbar (\phi_s-\phi_{ch})}  \right),  \phi_s-\phi_{ch}>\frac{E_g}{q}+\delta \\
\frac{2q}{h}  exp(- {\Lambda}  \frac{\pi}{2} \frac{1}{\hbar} \sqrt{m_{r}^*  E_g)} \left(\theta(|\frac{\phi_d-\phi_{ch}+E_g}{E_g}|)- \theta(|\frac{\phi_s-\phi_{ch}}{E_g}|) \right), else \\
\end{cases}}
\end{equation}
where $\alpha$ and $\theta$ are defined in Appendix \ref{App_C} and $\delta$ equals $2q/\alpha^2$ (the other parameters have been previously introduced).
It is clear that the compact model given by (\ref{eq:IV}) reproduces with good accuracy and in a continuous fashion the results obtained through NEGF both in the ON- and OFF-state of the device. The input parameters of the analytic approach are: $m^*_h$, $m^*_e$, $E_g$, $L_{ch}$, and $\Lambda$. The three former are material dependent quantities while the latter two are determined by the device design.  Notice that due to the low density-of-sates (DOS) of the channel material (InAs), the relation between channel-potential and gate-voltage is close to one-to-one. If a channel material with a high DOS is chosen, the model still provides the current vs. channel-potential characteristic but a capacitor network (series of oxide \cite{optK}, quantum \cite{Hesam1, WKB1, Cq1, Cq2}, and offset \cite{Coff} capacitances) is needed to translate the channel-potential into a gate-voltage. 

Therefore, the proposed analytic model enables the quantitative prediction and analysis of material and geometry dependence of the full I-V characteristics of \emph{high performance} TFETs in direct bandgap semiconductors.  
Notice that this analytic approach captures most of the relevant tunneling phenomena in TFETs without relying on arbitrary fitting parameters. Thus the model provides a simple and useful  alternative to exact numerical methods such as NEGF which have high computational burden.

\section{Conclusion}
In this work, new expressions have been introduced to describe two key parameters of the tunneling transmission in tunneling devices: the tunneling effective mass $m^*_t$ and the total band bending distance $\Lambda$. The equations for $m^*_t$ and $\Lambda$ have been provided in metal-semiconductor (SB tunneling) and p-n junctions (BTBT). It has been proven that the concept of tunneling effective mass not only captures both the ellipticity of the complex bands and the electron/hole dual behavior, but it can also be applied to unconventional semiconductors such as GNRs. The combination of $m^*_t$ and $\Lambda$ leads to a modified FN formula which is more accurate for the calculation of BTBT currents in the ON-state of TFETs. Moreover, a new equation has been provided to describe the OFF-state performance of TFETs while providing a continuous transition at the ON-OFF threshold using the continuity factor. Ultimately, the results of the new analytical model are shown to be in good agreement with full band quantum transport simulations both in the ON- and OFF-state of the TFET. The analytic equations reveal that both ON- and OFF-currents decay exponentially with $\sqrt{m^*_r E_g}$. Accordingly, the channel material of TFETs needs to have an optimum $\sqrt{m^*_r E_g}$ to satisfy the required $I_{OFF}$ and $I_{ON}$. 
\appendices
\section{BTBT effective mass}
\label{App_A}
An accurate calculation of band-to-band tunneling transmission requires considering the elliptic nature of the complex band structure (decaying states). Notice that the total transmission depends exponentially on $k_{im}(E)$, $T_{WKB}^{BTBT}=exp(-2/qF \int_0^{E_g}k_{im}dE)$. The complex wave-vector $k_{im}(E)$ has electron and hole like regions with transition energy of $E_q$. 
\begin{equation}
\label{eq:A1}
\underbrace{\int_0^{E_g}k_{im}(E)dE}_{\Omega}= \int_0^{E_q}k_{im}^h(E)dE+ \int_{E_q}^{E_g}k_{im}^e(E)dE
\end{equation}
Using the analytic equations for an elliptic complex band structure \cite{elliptic}, one can derive:
\begin{multline}
\label{eq:A2}
\Omega = \frac{1}{\hbar} \int_0^{E_q}\sqrt{2m_h^*  E(1-\frac{E}{2E_q})} dE+ \\ 
\frac{1}{\hbar} \int_{E_q}^{E_g}\sqrt{2m_e^* (E_g-E)\left(1-\frac{E_g-E}{2(E_g-E_q)} \right) } dE
\end{multline}
Evaluation of the integral leads to:
\begin{equation}
\label{eq:A3}
\Omega = \frac{\pi}{16} \frac{4}{\hbar} \sqrt{m_h^* E_q} E_q+ \frac{\pi}{16} \frac{4}{\hbar} \sqrt{m_e^* (E_g-E_q ) }(E_g-E_q) 
\end{equation}
\begin{equation}
\label{eq:A4}
\Omega =\frac{\pi}{4\hbar} \sqrt{m_r^*} E_g^{3/2}
\end{equation}

\section{Effective mass for Schottky barriers}
\label{App_B}
In this section $m^*_t$ is derived for a SB with a barrier height of $\phi_b$ for electrons. The elliptic complex band in the electron branch is described by \cite{elliptic}:
\begin{equation}
\label{eq:A5} 
k_{ellip}=\frac{1}{\hbar} \sqrt{2m_e^* (E_g-E)\left(1-\frac{E_g-E}{2(E_g-E_q)} \right) },~ E_q<E<E_g
\end{equation}
A parabolic approximation to (\ref{eq:A5}) is given by
\begin{equation}
\label{eq:A6} 
k_{para}=\frac{1}{\hbar} \sqrt{2m_e^* (E_g-E)\left(1-\frac{\omega}{2(E_g-E_q)} \right) }
\end{equation}
where $E_q=E_g \times m_e/(m_h+m_e)$ and $\omega$ is a constant (in units of energy). Following a procedure similar to that established in power-law/polynomial conversions \cite{poly}, an error function $Err$ is defined over the tunneling window through the Schottky-Barrier (i.e. energies from $E_g-\phi_b$ to $E_g$) as follows:
\begin{equation}
\label{eq:A7} 
Err = \int_{E_g-\phi_b}^{E_g} (k_{ellip} - k_{para})^2 dE
\end{equation}
The optimum $\omega$ with the minimum error can be obtained from
\begin{equation}
\label{eq:A8} 
\frac{\partial Err}{\partial \omega} = 0
\end{equation}
The optimum value of $\omega$ is a complicated function of $E_g$, $E_q$, and $\phi_b$. However, a polynomial expression for $\omega$ can be found either by Taylor expansion or the successive integration method \cite{SIM} leading to:
\begin{equation}
\label{eq:A9} 
\omega^{opt}= \frac{2}{3} \phi_b + \frac{1}{144} \frac{\phi_b^2}{2(E_g-E_q)}+...
\end{equation}
Taking the leading term in (\ref{eq:A9}) in combination with (\ref{eq:A6}) leads to the effective mass in a metal-semiconductor junction with barrier height $\phi_b$:
\begin{equation}
\label{eq:A10} 
m_t^*|_{SB} ~\approx m_e^* \left(1- \frac{\phi_b}{3(E_g-E_q)}\right) = m_e^* \left(1- \frac{\phi_b}{E_g} \frac{m_e^*}{3m_r^*}\right)        
\end{equation}
Equation (\ref{eq:A10}) can also be used for holes by replacing $m_e^*$ with $m_h^*$ and considering the hole Schottky barrier height (i.e. $E_g - \phi_b$). For the materials with $m_h^* \approx m_e^*$ or $m_h^* \gg m_e^*$, (\ref{eq:A10}) can be rewritten approximately as
\begin{equation}
\label{eq:A10_2} 
m_t^*|_{SB} ~\approx m_e^* \left(1- \frac{\phi_b}{E_g}\right) + 0.7 m_r^* \left(\frac{\phi_b}{E_g}\right)
\end{equation}

\section{Modeling the OFF-state}
\label{App_C}
\begin{figure}[!t]
        \centering
        \begin{subfigure}[b]{0.3\textwidth}
                \includegraphics[width=\textwidth]{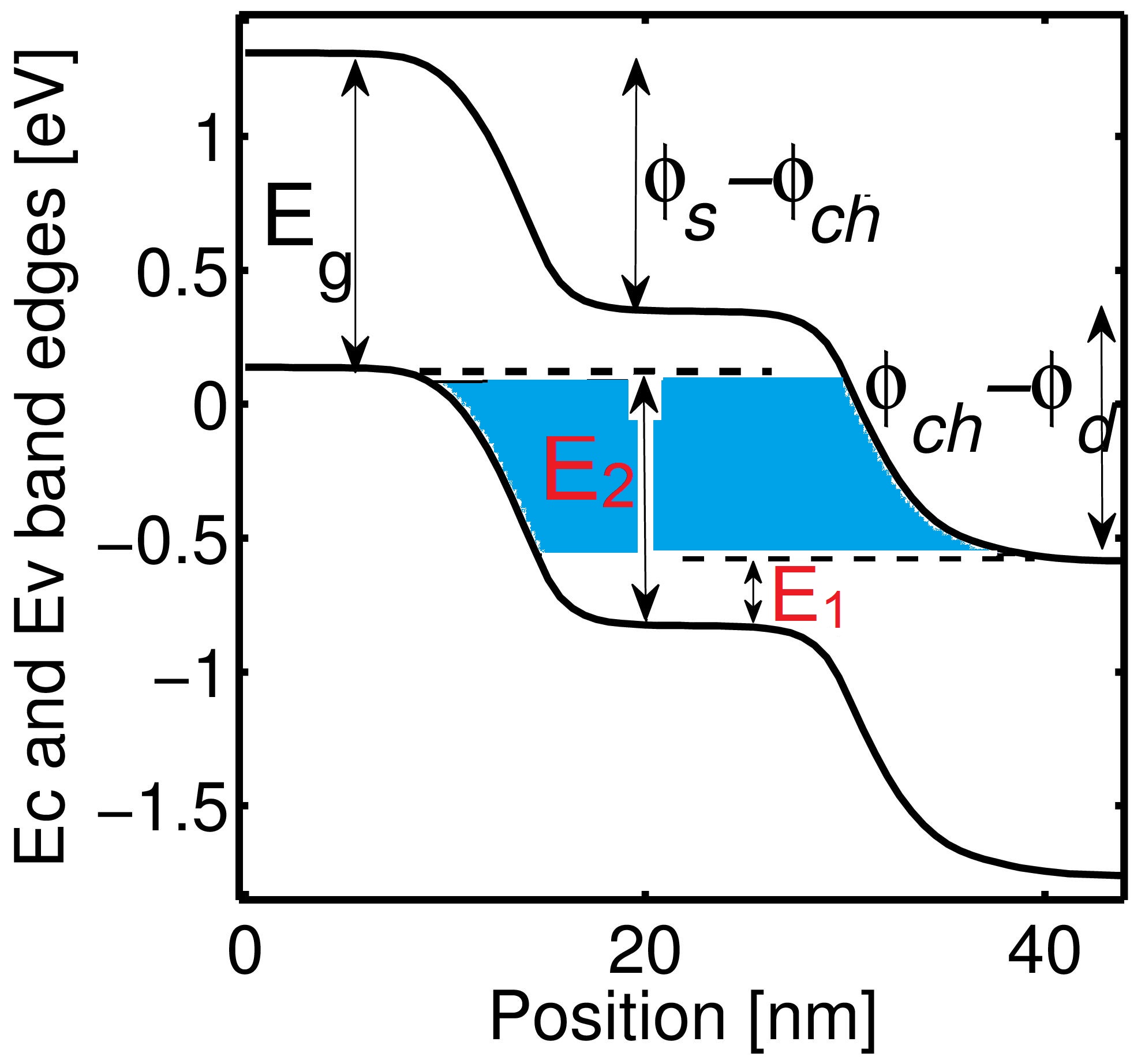}
                \label{fig:IdVg_dop}
        \end{subfigure}%
              \vspace{-1.5\baselineskip}                
        \caption{Band diagram and tunneling window (blue region) of a TFET in the OFF-state.}\label{fig:off_2}
\end{figure}

The direct tunneling between source and drain in the off-state of the device is significantly (but not completely) determined by a spatially constant $k_{im}(E)$. In this case the transmission can be written as:
\begin{multline}
\label{eq:App4}
T_{OFF}= \int_{E_1}^{E_2}exp\left(-2\int k_{im} (E)dx \right) dE \\
T_{OFF}\approx\int_{E_1}^{E_2}exp \left(-2k_{im}(E)L^' \right) dE
\end{multline}
\noindent where $L^'$ is the portion of the channel over which $k_{im}(E)$ is spatially constant. The energies $E_1$ and $E_2$ define the tunneling energy window in the off-state of the device as shown in Fig. \ref{fig:off_2}. The actual form of the transmission considering elliptic bands is simplified by assuming that the electron and hole effective masses are both equal:
\begin{equation}
\label{eq:A11}
m^*=m_e^*=m_h^*=2m_r^* 
\end{equation}
\begin{equation}
\label{eq:A12}
T_{OFF}=\int_{E_1}^{E_2}exp\left(- \frac{2L^'}{\hbar} \sqrt{m^*  E^'(1-E^'/E_g )}\right) dE^' 
\end{equation}
All the energies in (\ref{eq:A12}) are referenced with respect to the edge of the valence band ($\phi_{valence}=\phi_{ch}-E_g $) and therefore:
\begin{align}
\label{eq:A13}
E^'=E-\phi_{ch}+E_g  \nonumber \\
E_1=\phi_d-\phi_{ch}+E_g  \nonumber \\
E_2=\phi_s-\phi_{ch} 	
\end{align}

Making the change of variable  $x=E'/E_g$  leads to:
\begin{equation}
\label{eq:A16}
T(x_1,x_2,\alpha)=\int_{x_1}^{x_2}exp\left(-\alpha\sqrt{x(1-x)}\right) dx, 
\end{equation}
where 
\begin{align}
\label{eq:A17}
\alpha= \frac{2L^'}{\hbar} \sqrt{m^* E_g} \nonumber \\
x_1=E_1/E_g \nonumber \\
x_2=E_2/E_g 
\end{align}

The integral shown in (\ref{eq:A16}) does not have an analytic solution. However, since $\sqrt{x(1-x)}$ can be fitted to a polynomial function (\ref{eq:A16}) can be integrated analytically in an approximated manner. The polynomial approximation is given by:
\begin{equation}
\label{eq:A18}
\sqrt{x(1-x)}\approx 
\begin{cases}
\sqrt{x} &  x<0.1 \\
-1.2x^2+1.2x+0.2 &  0.1<x<0.9 \\
\sqrt{1-x} & 0.9<x \\
\end{cases}
\end{equation}
Using (\ref{eq:A18}), the total transmission is found to have an analytic solution:
\begin{equation}
\label{eq:A19}
T(x_1,x_2,\alpha) \approx \theta(x_2)-\theta(x_1)
\end{equation}
where the function $\theta(x)$ is defined as:
\begin{equation}
\label{eq:App4}
\theta(x)= G(x)+ 0.8 \frac{exp(-0.5\alpha)}{\sqrt{\alpha}}  erfi\left(0.55{\sqrt{\alpha} (-1+2x)}\right) 
\end{equation}
G(x) is due to the terms $\sqrt{x}$ and $\sqrt{1-x}$:
\begin{equation}
\label{eq:App4_2}
G(x)= 
\begin{cases}
g(x) &  x<0.1 \\
g(0.1) &  0.1<x<0.9 \\
g(1-x)+g(0.1) & 0.9<x \\
\end{cases}
\end{equation}
where $g(x)$ is 
\begin{equation}
\label{eq:A4_3}
g(x) = \frac{2}{\alpha^2} exp(-\alpha\sqrt{x})(-\alpha\sqrt{x}-1).
\end{equation}
In the case of $m^*_e \neq m^*_h$, the electron effective mass should be considered in (\ref{eq:A12}), $m^*=m^*_e$, for the n-branch of I-V.

\section*{Acknowledgment}
This work was supported in part by the Center for Low Energy Systems Technology (LEAST), one of six centers of STARnet, a Semiconductor Research Corporation program sponsored by MARCO and DARPA. 


\end{document}